\documentclass[aps,prd,preprint,groupedaddress,showpacs,longbibliography]{revtex4-1}

\usepackage{epsfig}
\usepackage{graphics}
\usepackage{slashed}
\usepackage{color}
\usepackage{multirow}
\begin{document}

\leftmargin -2cm
\def\choosen{\atopwithdelims..}

\boldmath
\title{Sudakov resummation from the BFKL evolution} \unboldmath

\author{\firstname{M.A.} \surname{Nefedov}} \email{nefedovma@gmail.com}
\affiliation{Universit\'{e} Paris-Saclay, CNRS, IJCLab, 91405 Orsay, France}
\affiliation{Samara National Research University, Moskovskoe Shosse,
34, 443086, Samara, Russia}

\begin{abstract}
The Leding-Logarithmic (LL) gluon Sudakov formfactor is derived from rapidity-ordered BFKL evolution with longitudinal-momentum conservation. This derivation further clarifies the relation between High-Energy and TMD-factorizations and can be extended beyond LL-approximation as well.   
\end{abstract}



\maketitle

\section{Introduction}
\label{sec:Int} 

The interplay between effects of High-Energy QCD and Transverse-Momentum Dependent (TMD) Factorization has become a topic of intensive theoretical studies in recent years, see e.g. Refs.~\cite{Balitsky:2015qba, Balitsky:2016dgz,Balitsky:2019ayf,Kotko:2015ura,Altinoluk:2019wyu,Altinoluk:2019fui,
Altinoluk:2020qet,Boussarie:2020fpb,Xiao:2018esv,Mueller:2015ael,Mueller:2013wwa}. Regions of applicability of these two formalisms overlap when observables related with transverse-momentum ($q_T$) are studied in the process containing another large scale $M$ in hadronic or lepton-hadron collision with large center-of-mass energy $\sqrt{S}$, while all three scales become hierarchical: $\sqrt{S}\gg M \gg q_T$.  The rapidity-evolution equation for gluon TMD Parton Distribution Function has been conjectured few years ago~\cite{Balitsky:2015qba, Balitsky:2016dgz}, which unifies the non-linear Balitsky-Kovchegov~\cite{Balitsky:1995ub, Kovchegov:1999yj} evolution equation in the small-$x$ limit with the linear Collins-Soper-Sterman(CSS)-type evolution of TMD PDFs familiar at moderate values of $x$ (see e.g. the monograph~\cite{CollinsQCD} for the review of TMD factorization). Unfortunately the systematic study of this evolution equation is a very challenging task. Its real-emission kernel is essentially three-dimensional, interpolating between small-$x$ (Regge) limit with characteristic non-trivial transverse-momentum dynamics and moderate-$x$ limit, for which longitudinal momentum conservation is crucial~\footnote{From this point of view, it is interesting to study the relation between real emission kernel of Refs.~\cite{Balitsky:2015qba, Balitsky:2016dgz} and TMD gluon-gluon splitting function derived in Ref.~\cite{Hentschinski:2017ayz}}. More recently, the origin of corrections $\sim (\alpha_s \ln^2 (M^2/q_T^2))^n$ -- the so-called ``Sudakov'' double logarithms, in the approach of Refs.~\cite{Balitsky:2015qba, Balitsky:2016dgz} had been further clarified in the Ref.~\cite{Balitsky:2019ayf} where the roles of conformally-invariant cut-off for rapidity divergences as well as of conservation of dominating light-cone component of momentum where emphasized. However, connection to the physics of standard Balitsky-Fadin-Kuraev-Lipatov (BFKL)~\cite{BFKL1,BFKL2,BFKL3} evolution equation, which is the cornerstone of High-Energy QCD, seems to be lost in these recent developments.

The goal of the present paper is to restore this connection. As it will be shown below, the Sudakov suppression of the gluon Unintegrated PDF(UPDF) in the limit $q_T\ll M$ can be reproduced starting from the standard Leading-Logarithmic BFKL equation if one imposes the conservation of leading light-cone momentum component through the cascade of real emissions. 

The present study has also been motivated by few other considerations. First, it is known, that resummation of higher-order corrections enhanced  by $\ln M/q_T$ in the formalisms of Soft-Collinear Effective Theory(SCET)~\cite{Becher:2011dz, Becher:2011xn, Chiu:2012ir} and TMD-factorization~\cite{CollinsQCD} are driven by rapidity divergences. On the other hand, the kernel of BFKL equation can be derived as a coefficient of rapidity-divergent part of four-Reggeon Green's function in Lipatov's gauge-invariant EFT for high-energy processes in QCD~\cite{Lipatov95}, see e.g. Ref.~\cite{Bartels:2012mq}. The gluon Regge trajectory at one~\cite{Chachamis:2012cc} and two loops~\cite{Chachamis:2012gh,Chachamis:2013hma} can be computed as the coefficient of rapidity divergence of Reggeized gluon self-energy. These results suggest a close connection between rapidity divergences in SCET and BFKL physics, see e.g. Ref.~\cite{Rothstein:2016bsq,Moult:2017xpp} for an approach to incorporate BFKL physics into SCET language.

Further motivation is, that in the phenomenology of High-Energy factorization it has long been understood, that the most useful phenomenological UPDFs must include effects of Sudakov resummation besides BFKL effects. A well-known example of such UPDF is the Kimber-Martin-Ryskin-Watt formula~\cite{Kimber:2001sc, Watt:2003mx, Watt:2003vf, Martin:2009ii}. Combination of Sudakov-improved UPDF with gauge-invariant matrix elements with Reggeized (off-shell) initial-state partons gives excellent phenomenological results for observables such as Dijet~\cite{Nefedov:dijet, Bury:forward-dijet, vanHameren:2020rqt}, multi-jet~\cite{Kutak:4-jet} correlations, angular correlations of pairs of $B$~\cite{Karpishkov:BB} and $D$-mesons~\cite{Maciula:DD, Maciula:2019izq}, $J/\psi$ pair production~\cite{He:di-Jpsi}, Drell-Yan process~\cite{Nefedov:2012cq, Nefedov:2018vyt, Nefedov:2020ugj} and many other observables. Derivation of Sudakov formfactor from BFKL evolution in which $t$-channel partons are Reggeized, provides further evidence, that such phenomenological approach is self-consistent.  
      
The present paper has the following structure: in the section~\ref{sec:HEF} the High-Energy Factorization is reviewed and the concept of Unintegrated PDF is introduced; in the Sec.~\ref{sec:cascade} the modified BFKL Green's function with longitudinal momentum conservation is computed; in the Sec.~\ref{sec:Sudakov} the Sudakov formfactor for UPDF is derived in the leading (doubly-)logarithmic approximation, in the Sec.~\ref{sec:Heur-der}, the same result is re-derived in a heuristic way to expose more clearly it's physical content and in the Sec.~\ref{sec:disc} relations of our derivation with other results in literature and directions for further studies are discussed. To the Appendix, we put the details of the derivation of the function $\Delta\chi(\gamma,\lambda)$, which is used for the study of the modified BFKL Green's function.

\section{High-energy factorization and unintegrated PDF}
\label{sec:HEF}
The high-energy factorization was originally introduced in Refs.~\cite{Catani:1990xk, Catani:1990eg, Collins:1991ty, Catani:1994sq} as a tool to resum higher-order QCD corrections $\sim\alpha^n_s\ln^n(1/z)$ in the hard-scattering coefficient of Collinear Parton Model. Here variable $z$ can be e.g. the ratio of the squared transverse mass ($M_T$) of the final-state, which is detected in a hard inclusive reaction of interest to the {\it partonic} center-of-mass energy $z=M_T^2/\hat{s}$. See e.g. Refs.~\cite{Ball:2017otu, Abdolmaleki:2018jln, Hautmann:2002tu, Harlander:2009my} and references therein for applications of this formalism to Deep-Inelastic Scattering and Higgs boson production. However, besides large logarithms $\ln(1/z)$ the ``Sudakov'' large logarithms $\ln (M_T/|{\bf p}_T|)$ also could be important if transverse-momentum of the final-state of interest $|{\bf p}_T|\ll M_T$. Therefore the following slightly more general form of High-Energy Factorization, incorporating to some extent both resummations, is more useful:  
\begin{eqnarray}
 d\sigma &=& \int\limits_{0}^1 \frac{dx_1}{x_1} \int \frac{d^2{\bf q}_{T1}}{\pi} \Phi_i(x_1,{\bf q}_{T1},\mu_Y) \times \nonumber \\
 && \int\limits_{0}^1 \frac{dx_2}{x_2} \int \frac{d^2{\bf q}_{T2}}{\pi} \Phi_j(x_2,{\bf q}_{T2},\mu_Y)\ d\hat{\sigma}_{ij}(x_1,{\bf q}_{T1},x_2,{\bf q}_{T2}), \label{eq:HEF} 
\end{eqnarray}
where indices $i,j=R,Q,\bar{Q}$ with $R$ denoting {\it Reggeized gluon} and $Q$(${\bar Q}$) are Reggeized (anti-)quarks, $\hat{\sigma}_{ij}$ is the HEF coefficient function, which can be obtained for any QCD hard process using Lipatov's High-Energy EFT~\cite{Lipatov95}. Four-momenta ($q_{1,2}^\mu$) of Reggeized partons are equal to $q_{1,2}^\mu=x_{1,2} P_{1,2}^\mu+q_{T1,2}^{\mu}$, where $P_{1,2}$ are four-momenta of colliding hadrons.  The {\it unintegrated PDFs} (UPDFs) $\Phi_i$ in Eq. (\ref{eq:HEF}) depend on the {\it rapidity scale} $\mu_Y\sim M_T$. The  definition of $\mu_Y$ is
\begin{equation}
\mu_Y =  x_1P_1^+e^{-Y_\mu} =  x_2P_2^-e^{Y_\mu}, \label{eq:muY-def} 
\end{equation}
where $Y_\mu$ is the rapidity of the hard final state of $d\hat{\sigma}$ and light-cone conponents are introduced in the center-of-mass frame of $P_{1,2}$ as $P_{1}^+=P_1^0+P_1^3$, $P_2^-=P_2^0-P_2^3$. The gluon UPDF in pure gluodynamics (QCD with $n_F=0$) will be main object of our present study. In the purely perturbative approach, ignoring both the non-perturbative content in the TMD ($|{\bf p}_T|\ll M_T$) region as well as violations of collinear factorization due to the genuine gluon-saturation effects, and in the Leading-Logarithmic approximation in BFKL-sense, which is detailed below, the UPDF is related with the collinear momentum-density PDF $\tilde{f}_g(x,\mu_F)=xf_g(x,\mu_F)$ as follows:
\begin{equation}
\Phi_R(x,{\bf q}_T^2,\mu_Y)=\int\limits_x^1 \frac{dz}{z} {\cal C} \left( z, {\bf q}_T^2, \mu_Y,\mu_F \right) \tilde{f}_g\left(\frac{x}{z},\mu_F \right), \label{eq:Phi-def}
\end{equation}
where ${\cal C}$ is the resummation factor and $\mu_F$ is the factorization scale. The $\mu_F$-dependence should cancel-out in Eq.~(\ref{eq:Phi-def}), if the DGLAP-evolution for PDFs is taken in an approximation consistent with the approximation for ${\cal C}$. 

\begin{figure}
\begin{center}
\includegraphics[width=0.75\textwidth]{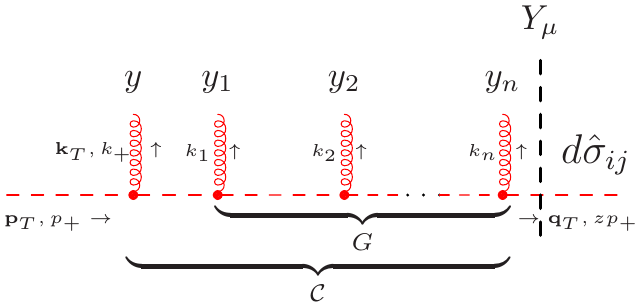}
\end{center}
\caption{Typical diagram for the amplitude corresponding to the resummation factor ${\cal C}$. The dashed line denotes Reggeized gluon and solid circles are Lipatov's vertices. The {\it rebounded gluon} has transverse momentum ${\bf k}_T$ and rapiditiy $y$. Rapidities of all emitted gluons are ordered: $y>y_1>y_2>\ldots>y_n>Y_\mu$. The gluon with momentum $p^+$ enters from collinear PDF, the non-zero ${\bf p}_T$ of this gluon is introduced to regularize collinear divergences (see the main text). The Reggeized gluon with transverse momentum ${\bf q}_T$ and $(+)$ momentum-component $zp_+$ enters into the coefficient function $d\hat{\sigma}_{ij}$ of Eq.~(\ref{eq:HEF}). \label{fig:ladder} }
\end{figure}

  For definiteness, let us talk about UPDF $\Phi_R(x_1,{\bf q}_{T1},\mu_Y)$ which is associated with partons flying in the positive rapidity direction. The factor ${\cal C}$ in this UPDF resums QCD corrections enhanced by rapidity gap $Y$ between the most forward gluon emission belonging to the hard-scattering coefficient of CPM, let us call it the {\it rebounded gluon}, and $Y_\mu$ of observed particles. The Leading Logarithmic Approximation for HEF resums terms $\sim (\alpha_s Y^2)^n$ and $\sim (\alpha_s Y)^n$ to all orders in $\alpha_s$ using LL BFKL evolution equation and such resummation is correct up to terms suppressed as $e^{-Y}$. The standard version of HEF~\cite{Collins:1991ty, Catani:1994sq} uses approximation $Y\sim \ln(1/z)$, which is too restrictive and misses a lot of interesting physics as we will see below. Instead we put $Y=y-Y_\mu$, where $y$ is the rapidity of the rebounded gluon and introduce the {\it collinearly un-subtracted} resummation factor as (see also the Fig.~\ref{fig:ladder} for more insight on on the notation):
  \begin{eqnarray}
  \tilde{\cal C}(z,{\bf q}_T^2,\mu_Y|{\bf p}_T^2) &=& \delta(z-1)\delta({\bf q}_T^2-{\bf p}_T^2) \nonumber \\&+& \hat{\alpha}_s\int\limits^{+\infty}_{Y_\mu} dy \int\frac{d^2{\bf k}_T}{\pi {\bf k}_T^2}\ G\Big({\bf q}_T^2 , z p_+ \Big\vert\ y-Y_\mu, ({\bf p}_T-{\bf k}_T)^2, p_+-k_+\Big),  \label{eq:C-unsub-def}
  \end{eqnarray}   
  where $\hat{\alpha}_s=\alpha_sC_A/\pi$, $G$ is the BFKL Green's function, $p_+=P_1^+x_1/z$ is the large (+) momentum component, entering the full CPM hard process from the collinear PDF and $k_+=|{\bf k}_T| e^y$ is the (+) light-cone momentum component of the rebounded gluon. The factor $\hat{\alpha}_s/{\bf k}_T^2$ is the square of Lipatov's vertex for emission of the rebounded gluon, and in Eq.~(\ref{eq:C-unsub-def}) we integrate over it's phase-space, parametrized by rapidity $y$ and transverse-momentum ${\bf k}_T$. Traditionally, see e.g. the textbook~\cite{kovchegov_levin_2012}, the BFKL Green's function depends only on the rapidity interval $Y$, and transverse-momenta of ``incoming'' (${\bf p}_T-{\bf k}_T$) and ``outgoing'' (${\bf q}_T$) Reggeized gluons. In our case, rapidity $Y$ does not uniquely define the longitudinal momentum of partons, and therefore $G$ additionally depends on the (+) momentum component of the $t$-channel parton entering ($p_+-k_+$) and leaving ($zp_+$) the Green's function. The evolution for $G$ with longitudinal-momentum dependence is described in the Sec.~\ref{sec:cascade}. 
  
  To regularize initial-state collinear divergences in $\tilde{\cal C}$, one introduces non-zero transverse momentum ${\bf p}_T$ for the incoming gluon in Eq.~(\ref{eq:C-unsub-def}). Usually the dimensional regularization is used for this purpose in higher-order calculations in CPM, but we find the regularization scheme with separate regulator for collinear divergences to be more convenient for BFKL-type calculations. Eventually we are interested in the CPM limit ${\bf p}_T\to 0$ for $\tilde{\cal C}$. If one passes to the Mellin space for $z$-dependence and ${\bf x}_T$-space for ${\bf q}_T$-dependence:
  \[
  \tilde{\cal C}(N,{\bf x}_T^2,\mu_Y|{\bf p}_T^2) = \int\limits_0^1 dz\ z^{N-1} \int d^2{\bf q}_T\ e^{i{\bf x}_T{\bf q}_T}\ \tilde{\cal C}(z,{\bf q}_T^2,\mu_Y|{\bf p}_T^2), 
  \]
  then for ${\bf p}_T\to 0$ collinear divergences will factorize as:
 \begin{eqnarray*}
&& \tilde{\cal C}(N,{\bf x}_T^2,\mu_Y|{\bf p}_T^2) = Z_{\rm coll.}(N,{\bf p}_T^2) {\cal C}(N,{\bf x}_T^2,\mu_Y,\mu_F),\\ 
&& Z_{\rm coll.}(N,{\bf p}_T^2) = \exp\left[ -\gamma_{gg}(N,\hat{\alpha}_s) \ln \frac{{\bf p}_T^2}{\mu_F^2} \right],   
 \end{eqnarray*}
  where $\gamma_{gg}(N,\hat{\alpha}_s)$ is the DGLAP anomalous dimension and ${\cal C}$ is the {\it collinearly-subtracted} resummation function we had been looking for. For example, if one is interested only in LLA resummation of corrections enhanced by $\ln (1/z)$, then one can ignore complications related with longitudinal momentum conservation and scale $\mu_Y$ and obtain the result:
  \begin{equation}
  {\cal C}_{{\rm LLA} \ln 1/z} (N,{\bf x}_T^2,\mu_F) = (1+O(\hat{\alpha}_s^3))\exp\left[ -\gamma_{gg}(N,\hat{\alpha}_s)\ln(\mu_F^2 \bar{\bf x}_T^2) \right], \label{eq:BFKL-UPDF}
  \end{equation}
  with $\bar{\bf x}_T={\bf x}_T e^{\gamma_E}/2$ and $\gamma_{gg}$ determined implicitly by the famous equation~\cite{Jaroszewicz:1982gr}:
  \[
  \frac{\hat{\alpha}_s}{N}\chi(\gamma_{gg}(N,\hat{\alpha}_s))=1,
  \]
  where
  \begin{equation}
  \chi(\gamma)=2\psi(1)-\psi(\gamma)-\psi(1-\gamma),\label{eq:Lipatov-chi}
  \end{equation}
  is Lipatov's characteristic function and $\psi(\gamma)=\Gamma'(\gamma)/\Gamma(\gamma)$. 
  
  However, collinear divergences arise from terms $\sim (\hat{\alpha}_s Y)^n$, while in the present paper we are interested in effects $\sim (\hat{\alpha}_s Y^2)^n$. With this accuracy, collinear divergences do not appear and $\tilde{\cal C}$ is finite in the limit ${\bf p}_T\to 0$ as it will be shown in Sec.~\ref{sec:Sudakov}.

\section{BFKL cascade with longitudinal momentum conservation}
\label{sec:cascade}

  In the order $O(\alpha_s^n)$ the BFKL Green's function contains up to $n$ emissions or real gluons with (+) momentum components $k_1^+,\ldots, k_n^+$, given by $k_i^+=|{\bf k}_{Ti}|e^{y_i}$ (see again the Fig.~\ref{fig:ladder}). The overall (+) momentum-component conservation in Eq. (\ref{eq:C-unsub-def}) can be expressed by the delta-function, which is factorisable via Fourier-transform w.r.t. $x_-$ coordinate:
  \[
  \delta \left( p_+-k_+ - k_1^+ -\ldots -k_n^+ - z p_+ \right) = \int\limits_{-\infty}^{+\infty} \frac{dx_-}{2\pi} e^{ix_- (p_+ (1-z)-k_+)} \prod\limits_{i=1}^n e^{-ix_-k^+_i} .
  \] 
  
  Therefore, to keep track of (+) momentum component conservation, it is enough to consider the $x_-$-dependent BFKL Green's function, which will satisfy the standard BFKL equation in rapidity space with $e^{-ix_-k_i^+}$-factor added to the real-emission term:
  \begin{eqnarray}
&&  G\Big({\bf q}_T^2 \Big\vert\ Y, {\bf p}_T^2, x_-\Big) = G_0\Big({\bf q}_T^2 \Big\vert\ {\bf p}_T^2, x_- \Big)+\int\limits_0^Y dy\ \Big\{2\omega_g({\bf p}_T^2) G\Big({\bf q}_T^2 \Big\vert\ y, {\bf p}_T^2, x_-\Big)  \nonumber \\
&&    + \hat{\alpha}_s  \int\frac{d^{2-2\epsilon}{\bf k}_T  }{\pi  (2\pi)^{-2\epsilon} {\bf k}_T^2 } \exp[-ix_- |{\bf k}_T| e^y] G\Big({\bf q}_T^2 \Big\vert\ y, ({\bf p}_T-{\bf k}_T)^2, x_-\Big) \Big\}, \label{eq:BFKL-int}
  \end{eqnarray}
  where we have introduced dimensional regularization with $\epsilon<0$ to regularize infra-red divergences, $\omega_g({\bf p}_T^2)$ is the one-loop Regge trajectory of a gluon:
  \begin{equation}
  \omega_g({\bf p}_T^2)=-\frac{\hat{\alpha}_s}{4} \int\frac{d^{2-2\epsilon} {\bf k}_T}{\pi (2\pi)^{-2\epsilon}} \frac{{\bf p}_T^2}{{\bf k}_T^2 ({\bf p}_T-{\bf k}_T)^2} = \frac{\hat{\alpha}_s}{2\epsilon} ({\bf p}_T^2)^{-\epsilon} \frac{(4\pi)^\epsilon \Gamma(1+\epsilon) \Gamma^2(1-\epsilon)}{\Gamma(1-2\epsilon)}, \label{eq:omega}
  \end{equation}
  and the initial condition $G_0$ is given by:
  \begin{equation}
  G_0\Big({\bf q}_T^2 \Big\vert\ {\bf p}_T^2, x_- \Big) = \int\limits_{0}^{+\infty}dq_+ e^{-i x_- q_+}\ \delta\left( \frac{zp_+}{q_+} - 1 \right) \delta({\bf q}_T^2-{\bf p}_T^2) = zp_+ e^{-iz x_-p_+}  \delta({\bf q}_T^2-{\bf p}_T^2). \label{eq:G0}
  \end{equation}
  
  Differentiating Eq.~(\ref{eq:BFKL-int}) w.r.t. $Y$ one obtains a differential form of BFKL equation: 
  \begin{equation}
  \frac{\partial G \Big({\bf q}_T^2 \Big\vert\ Y, {\bf p}_T^2, x_-\Big)}{\partial Y} = \hat{\alpha}_s \int d^{2-2\epsilon} {\bf k}_T\ K({\bf k}_T^2,{\bf p}_T^2, x_-, Y) G \Big({\bf q}_T^2 \Big\vert\ Y, ({\bf p}_T-{\bf k}_T)^2, x_-\Big),\label{eq:BFKL-diff-kT}
  \end{equation}
 with 
 \[
 K({\bf k}_T^2,{\bf p}_T^2, x_-, Y) = \delta^{(2-2\epsilon)}({\bf k}_T) \frac{({\bf p}_T^2)^{-\epsilon}}{\epsilon} \frac{(4\pi)^\epsilon \Gamma(1+\epsilon) \Gamma^2(1-\epsilon)}{\Gamma(1-2\epsilon)} + \frac{\exp[-ix_- |{\bf k}_T| e^Y]}{\pi (2\pi)^{-2\epsilon} {\bf k}_T^2}.
 \]
 
 We use the fact, that for ${\bf p}_T\neq 0$ the IR divergences cancel order-by-order in $\alpha_s$ in $G$, so one can safely introduce the Mellin transform for ${\bf p}_T^2$-dependence of the Green's function:
 \begin{equation}
 G \Big({\bf q}_T^2 \Big\vert\ Y, {\bf p}_T^2, x_-\Big) = \int\frac{d\gamma}{2\pi i} \frac{1}{{\bf p}_T^2} \left( \frac{{\bf p}_T^2}{\mu_Y^2} \right)^{\gamma} G \Big({\bf q}_T^2 \Big\vert\ Y, \gamma, x_-\Big),   \label{eq:G-Mellin}
 \end{equation}
 and the kernel acts on powers of ${\bf p}_T^2$ as:
 \begin{equation}
 \lim\limits_{\varepsilon\to 0} \int d^{2-2\epsilon} {\bf k}_T\ K({\bf k}_T^2,{\bf p}_T^2, x_-, Y) (({\bf p}_T - {\bf k}_T)^2)^{-1+\gamma} = ({\bf p}_T^2)^{-1+\gamma} \widetilde{\chi}(\gamma, x_-|{\bf p}_T| e^Y  ), \label{eq:chi-tilde-def}
 \end{equation}
 with the function $\widetilde{\chi}$ still depending on ${\bf p}_T$ through it's argument $\tilde{x}_- = x_-|{\bf p}_T| e^Y $. Explicitly, this function is:
 \begin{eqnarray}
 \widetilde{\chi}(\gamma,\tilde{x}_-)&=&\lim\limits_{\epsilon\to 0^-} \widetilde{\chi}(\gamma,\tilde{x}_-,\epsilon)\nonumber \\&=& \lim\limits_{\epsilon\to 0^-} \left[ \frac{1}{\epsilon} \frac{\Gamma(1+\epsilon)\Gamma^2(1-\epsilon)}{\Gamma(1-2\epsilon)} + \int\frac{d^{2-2\epsilon}{\bf l}_T}{\pi^{1-\epsilon}} \frac{\exp\left[-i\tilde{x}_- |{\bf l}_T| \right]}{{\bf l}_T^2 \left(({\bf n}_T-{\bf l}_T)^2\right)^{1-\gamma}} \right],\label{eq:chi-tilde-DEF} 
 \end{eqnarray}
 where ${\bf n}_T$ is an arbitrary fixed unit vector in transverse plane. One can write-down an alternative definition of $\widetilde{\chi}(\gamma,\tilde{x}_-)$ with infra-red divergence canceled point-by-point in phase-space  (see again the textbook~\cite{kovchegov_levin_2012}). Using the explicit form of the Regge-trajectory as an integral in transverse-momentum space (\ref{eq:omega}) and the identity:
 \[
 \frac{1}{{\bf l}_T^2 ({\bf n}_T-{\bf l}_T)^2} = \frac{1}{{\bf l}_T^2 \left[ {\bf l}_T^2 + ({\bf n}_T-{\bf l}_T)^2 \right]} + \frac{1}{({\bf n}_T-{\bf l}_T)^2 \left[ {\bf l}_T^2 + ({\bf n}_T-{\bf l}_T)^2 \right]}, 
 \] 
 one obtains:
 \begin{equation}
 \widetilde{\chi}(\gamma,\tilde{x}_-) = \int\frac{d^2{\bf l}_T}{\pi {\bf l}_T^2} \left[ \left(({\bf n}_T-{\bf l}_T)^2\right)^{\gamma-1} \exp\left[-i\tilde{x}_- |{\bf l}_T| \right] - \frac{1}{{\bf l}_T^2 + ({\bf n}_T-{\bf l}_T)^2}  \right]. \label{eq:chi-noeps}
 \end{equation}
 The latter definition is convenient for direct numerical evaluation of $\widetilde{\chi}(\gamma,\tilde{x}_-)$.
  However for analytic studies one would like to finally disentangle the ${\bf p}_T^2$-dependence from $\widetilde{\chi}(\gamma,\tilde{x}_-)$. To this end, it is necessary to introduce (inverse-)Mellin-transform over $\tilde{x}_-$ as follows, see the Appendix for detailed derivation:
  \begin{equation}
  \widetilde{\chi}(\gamma,\tilde{x}_-) = \chi(\gamma) + \int\limits_{C_\lambda} \frac{d\lambda}{2\pi i}\ \tilde{x}_-^\lambda \Delta\chi(\gamma,\lambda), \label{eq:chi-tilde-Mellin} 
  \end{equation}
  where $\chi(\gamma)$ is an ordinary BFKL characteristic function (\ref{eq:Lipatov-chi}), $\Delta\chi(\gamma,\lambda)$ is given by Eq.~(\ref{eq:delta-chi-gamma-lambda}) in the Appendix and the contour $C_\lambda$ passes {\it to the right} from singularity of $\Delta\chi(\gamma,\lambda)$ at $\lambda=0$ parallel and in the same direction with the imaginary axis, thus:
  \begin{equation}
  \lim\limits_{\tilde{x}_-\to 0} \widetilde{\chi}(\gamma,\tilde{x}_-) = \chi(\gamma), \label{eq:BFKL-limit} 
  \end{equation} 
  i.e. the limit $\tilde{x}_-\to 0$ corresponds to the usual BFKL evolution of the Green's function.   
 
 Using representation (\ref{eq:chi-tilde-Mellin}) and Eqns.~(\ref{eq:BFKL-diff-kT}), (\ref{eq:G-Mellin}) and (\ref{eq:chi-tilde-def}) one obtains the evolution equation for $G \Big({\bf q}_T^2 \Big\vert\ Y, \gamma, x_-\Big)$ at arbitrary $x_-$:
 \begin{eqnarray}
 \frac{\partial G \Big({\bf q}_T^2 \Big\vert\ Y, \gamma, x_-\Big)}{\partial Y} &=& \hat{\alpha}_s \chi(\gamma) G \Big({\bf q}_T^2 \Big\vert\ Y, \gamma, x_-\Big) \nonumber \\ &+&  2\hat{\alpha}_s \int\limits_{C_\lambda'}\frac{d\lambda}{2\pi i}\ (\mu_Y x_-e^Y)^{2(\gamma-\lambda)} \Delta\chi (\lambda, 2(\gamma-\lambda))  G \Big({\bf q}_T^2 \Big\vert\ Y, \lambda, x_-\Big), \label{eq:BFKL-chi-tilde} 
 \end{eqnarray} 
 where the contour $C_\lambda'$ goes in the positive direction of imaginary axis of $\lambda$ and lies {\it to the left} of singularity of $\Delta\chi(\lambda,2(\gamma-\lambda))$ at $\gamma=\lambda$ to be consistent with definition of the contour in Eq.~(\ref{eq:chi-tilde-Mellin}).
 
 Since one have to do the inverse Fourier transform over $x_-$ to finally obtain the UPDF via Eq.~(\ref{eq:C-unsub-def}), then both small and large values of $x_-$ will contribute. However equation (\ref{eq:BFKL-chi-tilde}) is difficult to solve for general values of $x_-$, because it is still an integro-differential equation. Due to Eq.~(\ref{eq:BFKL-limit}), the limit of small-$x_-$ corresponds to ordinary BFKL-evolution and reproduces the standard BFKL UPDF (\ref{eq:BFKL-UPDF}). In the present paper we will concentrate on the opposite limit of large $x_-$. This limit is controlled by singularities of $\Delta\chi(\lambda,2(\gamma-\lambda))$ to the right of integration contour in Eq.~(\ref{eq:BFKL-chi-tilde}) and hence one should look for the pole at $\gamma-\lambda=0$ to obtain the leading-power contribution.
 
  The detailed derivation in the Appendix shows, that function $\Delta\chi(\gamma,\lambda)$ indeed has a pole at $\lambda=0$:
 \begin{equation}
 \Delta\chi(\gamma,\lambda) = -\frac{2}{\lambda^2} - \frac{1}{\lambda} \left( 2\gamma_E + i\pi +\chi(\gamma) \right) + O(1). \label{eq:chi-tilde-exp}
 \end{equation}
 
 Before proceeding, it is instructive to understand the source of the $-2/\lambda^2$ pole. From Eq. (\ref{eq:chi-noeps}) one easily deduces the contribution of the region $|{\bf l}_T|\ll |{\bf n}_T|=1$ to $\widetilde{\chi}(\gamma,\tilde{x}_-)$ at $\tilde{x}_-\gg 1$:
 \begin{eqnarray*}
 \widetilde{\chi}(\gamma,\tilde{x}_-) &\simeq& \int\limits_0^{1}\frac{d{\bf l}^2_T}{{\bf l}_T^2}  \left[ \exp\left( - i\tilde{x}_- |{\bf l}_T| \right) - 1 \right]  \\
 &=& -2\ln\tilde{x}_- -2\gamma_E - i\pi +O(\tilde{x}_-^{-1}),
 \end{eqnarray*}
and the term $-2\ln\tilde{x}_-$ corresponds to $-2/\lambda^2$ pole in $\lambda$-plane. From this short calculation one can see, that double pole at $\lambda=0$ has infra-red origin and both real-emission and Regge-trajectory terms of the BFKL kernel contribute to it. 

Substituting the Eq. (\ref{eq:chi-tilde-exp}) into (\ref{eq:BFKL-chi-tilde}) and taking the residue at $\lambda=\gamma$ one obtains:
\begin{eqnarray}
\hspace{-0.5cm}\left[\frac{\partial}{\partial Y} - \hat{\alpha}_s \frac{\partial}{\partial \gamma}\right] G \Big({\bf q}_T^2 \Big\vert\ Y, \gamma, x_-\Big) &=& -\hat{\alpha}_s G \Big({\bf q}_T^2 \Big\vert\ Y, \gamma, x_-\Big) \left[ 2\gamma_E + i\pi + 2 Y + 2\ln (\mu_Y x_-)  \right]. \label{eq:BFKL-la-0}  
\end{eqnarray}
Note that characteristic function $\chi(\gamma)$ has cancelled. This reflects a lack of non-trivial transverse-momentum dynamics in the Sudakov limit -- a puzzling feature of this regime from the point of view of a physicist with BFKL background. 

The solution of this equation, satisfying boundary condition (\ref{eq:G0}) transferred from ${\bf p}_T^2$ to $\gamma$-space is:
\begin{eqnarray}
\hspace{-0.5cm} G \Big({\bf q}_T^2 \Big\vert\ Y, \gamma, x_-\Big) &=& zp_+ e^{-iz x_- p_+} \left( \frac{{\bf q}_T^2}{\mu_Y^2} \right)^{-\gamma} \exp\left[ -\hat{\alpha}_sY\left(2\gamma_E+i\pi+\ln({\bf q}_T^2 x_-^2)\right) - \hat{\alpha}_s Y^2 \right]. \label{eq:G-sol-full}
\end{eqnarray}

In the present study we will consider only leading effects due to $\hat{\alpha}_s Y^2$-term in the exponent, therefore in the next section we will use the simplified solution without subleading effects:
\begin{equation}
G \Big({\bf q}_T^2 \Big\vert\ Y, \gamma, x_-\Big) = zp_+ e^{-iz x_- p_+} \left( \frac{{\bf q}_T^2}{\mu_Y^2} \right)^{-\gamma} e^{-\hat{\alpha}_s Y^2}. \label{eq:G-sol-DL}
\end{equation}

\section{Sudakov formfactor in the LLA} 
\label{sec:Sudakov}

Substituting the solution (\ref{eq:G-sol-DL}) as well as inverse Mellin transform over $\gamma$ and Fourier transform over $x_-$ into Eq.~\ref{eq:C-unsub-def}, one obtains:
\begin{eqnarray*}
&&\tilde{\cal C}(z,{\bf q}_T^2,\mu_Y | {\bf p}_T^2) = \delta(z-1)\delta({\bf q}_T^2-{\bf p}_T^2)  \nonumber \\
&&+ \hat{\alpha}_s \int\frac{d\gamma}{2\pi i} \int\limits_0^\infty dY \int\frac{d{\bf k}_T^2}{{\bf k}_T^2} \left[ \int\limits_{-\infty}^{+\infty}\frac{dx_-}{2\pi} (zp_+) \exp \left(ix_-(p_+(1-z) - |{\bf k}_T|e^{Y_\mu + Y}) \right) \right]  \nonumber \\
&& \times \frac{1}{({\bf p}_T-{\bf k}_T)^2} \left( \frac{({\bf p}_T-{\bf k}_T)^2}{{\bf q}_T^2} \right)^\gamma e^{-\hat{\alpha}_s Y^2},
\end{eqnarray*} 
where we have substituted $y=Y_\mu+Y$. Integral over $x_-$ leads to the delta-function, which can be used to integrate-out ${\bf k}_T^2=p_+^2(1-z)^2 e^{-2(Y_\mu+Y)}= \mu_Y^2 (1-z)^2 e^{-2Y}/z^2$, where we have used the relation $Y_\mu=\ln (zp_+)/\mu_Y$. Due to doubly-logarithmic exponent, the rapidity integral will converge, and therefore one can safely put ${\bf p}_T=0$ at this stage. Finally one gets:
 \begin{eqnarray*}
\tilde{\cal C}(z,{\bf q}_T^2,\mu_Y | {\bf p}_T^2=0) = \delta(z-1)\delta({\bf q}_T^2)  + \hat{\alpha}_s \frac{2z^3}{\mu_Y^2 (1-z)^3} \int\frac{d\gamma}{2\pi i} \left( \frac{\mu_Y^2}{{\bf q}_T^2} \frac{(1-z)^2}{z^2} \right)^\gamma  J(\gamma,\hat{\alpha}_s),
\end{eqnarray*} 
where
\[
J(\gamma,\alpha)=\int\limits_0^\infty dY \exp\left[ -\alpha Y^2 + 2Y(1-\gamma) \right] = \frac{\sqrt{\pi}}{2\sqrt{\alpha}}e^{\frac{(1-\gamma)^2}{\alpha}} \left[ 1 + {\rm Erf} \left( \frac{1-\gamma}{\sqrt{\alpha}} \right) \right]. 
\]

The latter function has the following expansion around $\alpha=0$:
\[
J(\gamma,\alpha)=\sqrt{\frac{\pi}{\alpha}} e^{\frac{(1-\gamma)^2}{\alpha}}+\frac{1}{2(\gamma-1)}+\sum\limits_{n=1}^\infty \frac{(2n-1)!}{2^{2n}(n-1)!} \frac{(-1)^n \alpha^n}{(\gamma-1)^{2n+1}},
\]
so moving the Mellin contour in $\gamma$-plane to the right one picks-up the perturbative contribution of the essential singularity at $\gamma=1$, thus obtaining:
 \begin{eqnarray}
&&\tilde{\cal C}(z,{\bf q}_T^2,\mu_Y | {\bf p}_T^2=0) = \delta(z-1)\delta({\bf q}_T^2) + \frac{\hat{\alpha}_s}{{\bf q}_T^2} \frac{z}{1-z} \exp 
\left[ -\frac{\hat{\alpha}_s}{4} \ln^2 \left( \frac{\mu_Y^2}{{\bf q}_T^2} \frac{(1-z)^2}{z^2} \right)  \right]. \label{eq:Sudakov-result}
\end{eqnarray} 

We conclude, that by picking-up the leading $\hat{\alpha}_s Y^2$-term in the exponent of BFKL Green's function (\ref{eq:G-sol-full}) we have reproduced the well-known ``Sudakov'' doubly-logarithmic suppression for gluon Unintegrated (or TMD) PDF at ${\bf q}_T^2\ll \mu_Y^2$~\cite{CollinsQCD, Dokshitzer:1978hw}. The Eq.~(\ref{eq:Sudakov-result}) also features doubly-logarithmic supression at $z\ll 1$, significance of which requires further analysis. The second term in this equation has a subleading power in $z$, due to overall factor of $z$ in front, while the standard BFKL result for UPDF (\ref{eq:BFKL-UPDF}) is of leading-power in $z$. The Eq.~(\ref{eq:BFKL-UPDF}) arises from consideration of small-$x_-\ll\mu_Y^{-1}$, i.e. BFKL limit of the solution of equation (\ref{eq:BFKL-chi-tilde}), while NLL terms in Sudakov regime arise from $\hat{\alpha}_s Y$-terms in the solution (\ref{eq:G-sol-full}), which is valid at $x_-\gg \mu_Y^{-1}$.

\section{Heuristic derivation}
\label{sec:Heur-der}

\begin{figure}
\begin{center}
\includegraphics[width=0.75\textwidth]{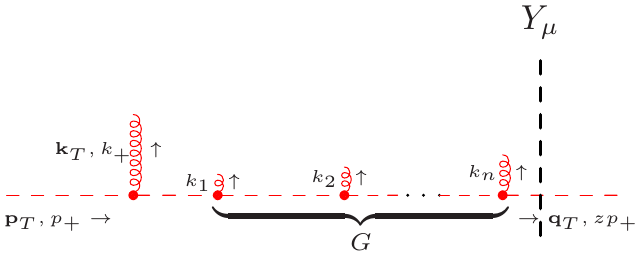}
\end{center}
\caption{ The Sudakov cascade. Length of gluon lines designates magnitude of their transverse momenta. Rapidities of all emitted gluons are ordered: $y>y_1>y_2>\ldots>y_n>Y_\mu$. \label{fig:Sud-cascade} }
\end{figure}

More intuitive derivation of the LL result for the Green's function (\ref{eq:G-sol-DL}) can be given as follows. From the lack of non-trivial $x_-$ or $\gamma$-dependence of the doubly-logarithmic solution (\ref{eq:G-sol-DL}) one concludes, that in this approximation, transverse and longitudinal momenta of all of the real emissions on the stage of evolution of the Green's function are negligible in comparison to $|{\bf q}_T|$ and $\mu_Y$ respectively (Fig.~\ref{fig:Sud-cascade}). In the regime $|{\bf q}_T|\ll \mu_Y$, both of this conditions are satisfied by the following cut on $(+)$- component of real-emission momentum:
\[
k_i^+=|{\bf k}_{Ti}| e^{y_i} \ll |{\bf q}_T|,
\]
where with logarithmic accuracy one can replace $\ll\to <$ to obtain the following upper limit for the transverse momentum of real emission:
\begin{equation}
|{\bf k}_{Ti}| < |{\bf q}_T| e^{-y_i}. \label{eq:Sudakov-cut}
\end{equation}
This special kinematic situation is represented schematically in the Fig.~\ref{fig:Sud-cascade} and can be called -- {\it the Sudakov cascade} of emissions. 

The cut (\ref{eq:Sudakov-cut}) can be introduced to the well-known iterative solution of BFKL equation (see e.g. Ref.~\cite{Schmidt:1996fg}) with explicit Regge-factors $\exp\left[ 2\omega_g({\bf q}_{Ti}) (y_{i+1}-y_{i}) \right]$ for each pair of adjacent emssions:
\begin{eqnarray}
G \Big({\bf q}_T^2 \Big\vert\ Y, {\bf p}_T^2 \Big) &=&  \sum\limits_{n=0}^{\infty} \frac{\hat{\alpha}_s^n}{\left[\pi (2\pi)^{-2\epsilon}\right]^n} \int \frac{d^{2-2\epsilon}{\bf k}_{T1}}{{\bf k}_{T1}^2}\ldots \frac{d^{2-2\epsilon}{\bf k}_{Tn}}{{\bf k}_{Tn}^2}\ \delta\left( {\bf p}_T-{\bf k}_{T1}-\ldots-{\bf k}_{Tn} - {\bf q}_T \right) \nonumber \\ 
&\times& \int \limits_0^Y dy_1\ e^{2\omega({\bf p}_{T}) y_1} \int\limits_{ y_1}^Y dy_2\ e^{2\omega({\bf q}_{T1}) (y_2-y_1)}\ldots \int\limits_{ y_{n-1}}^Y dy_n\ e^{2\omega({\bf q}_{Tn}) (Y-y_{n-1})},  \label{eq:G-iter}
\end{eqnarray}
where ${\bf q}_{Ti}={\bf p}_T-{\bf k}_{T1}-\ldots - {\bf k}_{Ti}$.

Applying the cut (\ref{eq:Sudakov-cut}) to Eq.~(\ref{eq:G-iter}) one obtains:
\begin{eqnarray}
G \Big({\bf q}_T^2 \Big\vert\ Y, {\bf p}_T^2 \Big) &\simeq& \delta({\bf p}_T-{\bf q}_T) e^{2\omega_g({\bf q}_T) Y}\sum\limits_{n=0}^{\infty} \hat{\alpha}_s^n \int \limits_0^Y dy_1\ \ldots \int \limits_{y_{n-1}}^Y dy_n\ J_\perp(y_1) \ldots J_\perp(y_n), \label{eq:G-iter-appr}
\end{eqnarray}
with
\[
J_\perp(y)=\int\frac{d^{2-2\epsilon}{\bf k}_T }{\pi (2\pi)^{-2\epsilon}  {\bf k}_T^2} \theta(|{\bf q}_T|e^{-y}-|{\bf k}_T|) = -\frac{1}{\epsilon} \frac{(4\pi)^\epsilon}{\Gamma(1-\epsilon)} \left( {\bf q}_T^2 e^{-2y}\right)^{-\epsilon}=L_{\epsilon}-2y+O(\epsilon),
\]
where $L_{\epsilon}=-1/\epsilon+\ln{\bf q}_T^2+\gamma_E-\ln 4\pi$, and one takes into account, that the transverse momentum of each emission is negligible in comparison to ${\bf q}_T$, so that all $t$-channel transverse momenta are equal ${\bf q}_{T1}=\ldots={\bf q}_{Tn}={\bf p}_T={\bf q}_T$ and all Regge-factors cancel-out, except one: $\exp\left[2\omega_g({\bf q}_T) Y\right]$. 

  Up to $O(\epsilon)$, the sum in Eq.~(\ref{eq:G-iter-appr}) leads to an exponent $\exp\left[ \hat{\alpha}_s Y (L_\epsilon-Y) \right]$, because:
  \[
  \int\limits_0^Y dy_1\ (L_\epsilon-2y_1) \int\limits_{y_1}^Y dy_2\  (L_\epsilon-2y_2) \ldots \int\limits_{y_{n-1}}^Y dy_n\  (L_\epsilon-2y_n) = \frac{1}{n!} \left[Y(L_\epsilon-Y)\right]^n,
\]
and infra-red divergence in $L_\epsilon$ cancels with the same divergence in the overall Regge factor, thus one obtains:
\[
G \Big({\bf q}_T^2 \Big\vert\ Y, {\bf p}_T^2 \Big) \simeq \delta({\bf p}_T-{\bf q}_T)e^{-\hat{\alpha}_sY^2},
\]
which coincides with Eq.~(\ref{eq:G-sol-DL}). From Eq.~(\ref{eq:G-sol-DL}) the final result for doubly-logarithmic Sudakov formfactor (\ref{eq:Sudakov-result}) follows as described in the Sec.~\ref{sec:Sudakov}. 

 In simple terms: the BFKL evolution has a natural tendency to produce emissions with substantial transverse and longitudinal momentum into a rapidity gap between rebounded gluon and hard process, thus the probability of ``Sudakov cascade'' configurations with no such emissions is suppressed by factor (\ref{eq:G-sol-DL}).
 
  The physical argument above also exposes the relation between Sudakov formfactor in our calculation and Sudakov formfactor in the Catani-Ciafaloni-Fiorani-Marcesini~\cite{Ciafaloni:1987ur,Catani:1989sg,Catani:1989yc,Marchesini:1994wr} and closely related to it Parton-Branching evolution equations~\cite{Martinez:2018jxt,Hautmann:2019biw,Martinez:2021chk}. In this approaches, the formfactor appears as probability of no emissions between two ``resolved'' splittings. The angular ordering criterion for resolution of the splitting essentially reduces to rapidity ordering, for small $|{\bf k}_T|\ll \mu_Y$ emissions. Thus an alternative formulation of above-mentioned evolution equations is possible, without explicit Sudakov formfactor, but in a form of BFKL-like evolution with longitudinal momentum conservation. We leave this interesting subject for future studies.

\section{Discussion and conclusions}
\label{sec:disc} 

  A few concluding remarks, connecting our result with other studies in the literature are in order. First, it is important to clarify the role of conformal symmeytry, which was an objective of Ref.~\cite{Balitsky:2019ayf}. One observes, that if one has the solution $ G \Big({\bf q}_T^2 \Big\vert\ Y, {\bf p}_T^2, x_-\Big)$ of the Eq.~(\ref{eq:BFKL-diff-kT}), then $ G \Big({\bf q}_T^2/\Lambda^2 \Big\vert\ Y, {\bf p}_T^2/\Lambda^2, \Lambda x_-\Big)$ with arbitrary re-scaling parameter $\Lambda$ is also a solution. This re-scaling symmetry may help to find a more convenient basis of functions for decomposition of the solution, than Mellin representation (\ref{eq:G-Mellin}) used in the present paper.
  
  The appearance of variable $x_-$ in our calculation, and separation of $x_-\ll\mu_Y^{-1}$ (BFKL) and $x_-\gg \mu_Y^{-1}$ (Sudakov) regions clearly resembles the ``thick shock-wave'' picture of Refs.~\cite{Balitsky:2015qba, Balitsky:2016dgz,Balitsky:2019ayf} and shows, that from point of view of rapidity-factorization, the Sudakov effects are the subset of sub-eikonal effects, related with longitudinal-momentum conservation and longitudinal structure of the target fields. It is interesting to connect this observations with recent developments in a field of sub-eikonal corrections in the High-Energy limit, see e.g. Refs.~\cite{Chirilli:2018kkw, Agostini:2019avp} and references therein.   
  
  The result (\ref{eq:Sudakov-result}) is also in agreement with results of Ref.~\cite{Mueller:2013wwa}, where consistency of small-$x$ resummation with the Sudakov resummation in ${\bf x}_T$-space has been verified in one-loop approximation on examples of Higgs hadro-production and heavy-quark photo-production. For the case of Higgs hadro-production, the term $-\alpha_s C_A/(2\pi)\ln M_H^2{\bf x}_T^2$ was found at one loop, while for the case of heavy-quark hadro-production, the coefficient of doubly-logarithmic term is $-\alpha_s C_A/(4\pi)$ as in our Eq.~(\ref{eq:Sudakov-result}). The ${\bf x}_T$-space is convenient, because transverse-momentum convolution of two unintegrated PDFs turns into a product of their ${\bf x}_T$-images. Since for the Higgs hadro-production one has two unintegrated PDFs for the initial state, the overall Sudakov formfactor in ${\bf x}_T$-space is given by the product of two such formfactors from each of the UPDFs. Hence the coefficient of double-logarithm in ${\bf x}_T$-space is two times larger for the Higgs case than for the photoproduction case, which has only one UPDF in the initial state. Unfortunately, already at NLL level the structure of Sudakov logarithms for production of colored final-states becomes highly process-dependent and can not be attributed solely to the initial-state Unintegrated or TMD PDFs. So HEF can not reproduce all subleading logarithms in the Sudakov limit.    
  
  The derivation presented above brings a hope to bridge the formal gap between High-Energy QCD and TMD factorization by deriving some results of the latter from BFKL physics. This is a mathematically challenging task, because the solution of Eq.~(\ref{eq:BFKL-chi-tilde}), which is valid in the whole range of $x_-$ have to be employed. But one might hope, that as it was in the case of collinear anomalous dimension~\cite{Jaroszewicz:1982gr}, the BFKL derivation might provide some constraints for higher-order structure of rapidity anomalous dimension, which governs rapidity evolution of TMD PDFs, as well as for the structure of $\ln 1/z$-enchenced corrections to collinear matching functions, which where renently found in the N$^3$LO calculation of Ref.~\cite{Luo:2019szz}. Another possible line of applications lies in the development of evolution equations based on the TMD splitting functions discovered in Refs.~\cite{Hentschinski:2017ayz, Gituliar:2015agu}.

\section*{Acknowledgements}

Author acknowledges Ian Balitsky, Giovanni A. Chirilli, Francesco Hautmann, Jean-Philippe Lansberg and other participants of workshops of the series ``Resummation Evolution Factorization'' and ``Quarkonia as Tools'' for useful discussions which have lead to the idea of this paper, as well as helpful referees for their comments. The work has been supported in parts by the Ministry of Education and Science of Russia via State assignment to educational and research institutions under project FSSS-2020-0014. The visit of M.A.N. to the IJClab has been supported by funding from the following sources:
\begin{itemize}
\item European Union’s Horizon 2020
research and innovation programme under the grant agreement No.824093 in order to contribute to the EU Virtual Access ``NLOAccess'',
\item the French ANR under the grant ANR-20-CE31-0015 (``PrecisOnium''),
\item the Paris-Saclay U. via the P2I Department of the Physics Graduate School,
\item the French CNRS via the ``GDR QCD'', via
the IN2P3 project ``GLUE@NLO'' and via the  IEA ``GlueGraph" (grant
agreement No.205210), 
\item the P2IO Labex via the Gluodynamics project.
\end{itemize}

\section*{Appendix: the characteristic function}

The function $\Delta\chi(\gamma,\lambda)$ in Eq.~(\ref{eq:chi-tilde-Mellin}) can be obtained from the $\epsilon\to 0$ limit of $\Delta\chi(\gamma,\lambda,\epsilon)$, which is given by the Mellin-transform of the real-emission term of the Eq.~(\ref{eq:chi-tilde-DEF}):
\begin{equation}
\Delta\chi(\gamma,\lambda,\epsilon)=\int\limits_0^\infty d\tilde{x}_-\ \tilde{x}_-^{-1-\lambda}  \int\frac{d^{2-2\epsilon}{\bf l}_T}{\pi^{1-\epsilon}} \frac{\exp\left[-i\tilde{x}_- |{\bf l}_T| \right]}{{\bf l}_T^2 \left(({\bf n}_T-{\bf l}_T)^2\right)^{1-\gamma}} .
\end{equation} 
After switching the order of integrations over $\tilde{x}_-$ and ${\bf l}_T$, integration over $\tilde{x}_-$ leads to a $\Gamma$-function and ${\bf l}_T$ can be integrated-out with the help of the formula
\[
 \int\frac{d^{2-2\epsilon}{\bf k}_T}{({\bf k}_T^2)^a (({\bf p}_T-{\bf k}_T)^2)^b} = \frac{\pi^{1-\epsilon} ({\bf p}_T^2)^{1-a-b-\epsilon} \Gamma(1-a-\epsilon)\Gamma(1-b-\epsilon)\Gamma(a+b+\epsilon-1) }{\Gamma(a)\Gamma(b)\Gamma(2-a-b-2\epsilon)},
\]
thus one obtains:
\begin{equation}
\Delta\chi(\gamma,\lambda,\epsilon)=e^{\frac{i\pi}{2}\lambda} \frac{\Gamma(-\lambda) \Gamma\left( \frac{\lambda}{2}-\epsilon \right) \Gamma(\gamma-\epsilon) \Gamma\left(1-\frac{\lambda}{2}-\gamma+\epsilon \right)}{\Gamma\left(1-\frac{\lambda}{2} \right) \Gamma(1-\gamma) \Gamma\left( \frac{\lambda}{2}+\gamma-2\epsilon \right)}. \label{eq:chi-gamma-lambda-eps}
\end{equation}

At this stage, the integration contour in the $\lambda$-plane passes {\it in between} the leftmost pole of $\Gamma(-\lambda)$ at $\lambda=0$, and the rightmost pole of $\Gamma\left( \frac{\lambda}{2}-\epsilon \right)$ at $\lambda=-2\epsilon$. In the limit $\epsilon\to 0^-$ this two poles pinch the contour to produce the infra-red divergence of the real-emission term. To extract this divergence explicitly, one moves the contour to the right of the pole at $\lambda=0$, thus introducing the contour $C_\lambda$ used in Eq.~(\ref{eq:chi-tilde-Mellin}) and picking-up the residue at this pole. The sum of the residue of $\Delta\chi(\gamma,\lambda,\epsilon)$ at $\lambda=0$ and the Regge-trajectory contribution in Eq.~(\ref{eq:chi-tilde-DEF}) gives the $\chi(\gamma)$-term in the Eq.~(\ref{eq:chi-tilde-Mellin}):
\[
\lim\limits_{\epsilon\to 0^-}\left[ -\mathop{\rm res}\limits_{\lambda=0} \Delta\chi(\gamma,\lambda,\epsilon) + \frac{1}{\epsilon} \frac{\Gamma(1+\epsilon)\Gamma^2(1-\epsilon)}{\Gamma(1-2\epsilon)} \right] = \chi(\gamma).
\] 

After moving the contour past the $\lambda=0$ pole, there is no pinch any more and one can safely put $\epsilon=0$, leading to the $\Delta\chi(\gamma,\lambda)$-function to be used in the Eq.~(\ref{eq:chi-tilde-Mellin}):
\begin{equation}
\Delta\chi(\gamma,\lambda)=\lim\limits_{\epsilon\to 0^-} \Delta\chi(\gamma,\lambda,\epsilon) = e^{\frac{i\pi}{2}\lambda} \frac{\Gamma(-\lambda) \Gamma\left( \frac{\lambda}{2} \right) \Gamma(\gamma) \Gamma\left(1-\frac{\lambda}{2}-\gamma \right)}{\Gamma\left(1-\frac{\lambda}{2} \right) \Gamma(1-\gamma) \Gamma\left( \frac{\lambda}{2}+\gamma \right)}. \label{eq:delta-chi-gamma-lambda}
\end{equation}
Expansion of this function around $\lambda=0$ is given in the Eq.~(\ref{eq:chi-tilde-exp}).

  Summing-up residues in the poles at $\lambda=n$ with integer $n>0$ and $\lambda=2(2n+2-\gamma)$ with $n=0,1,2,\ldots$ in Eq.~(\ref{eq:chi-tilde-Mellin}) one obtains the following exact expression for $\tilde{x}_-$-dependent characteristic function in terms of known hypergeometric functions:
  \begin{eqnarray}
  \tilde{\chi}_-(\gamma,\tilde{x}_-)&=&\chi(\gamma) - i\tilde{x}_-\frac{\Gamma\left(\frac{1}{2}-\gamma\right) \Gamma(\gamma)}{\Gamma(1-\gamma)\Gamma\left( \frac{1}{2}+\gamma\right)} {}_{1/2}F_{1/2}\left(\frac{3}{2},\frac{1}{2}+\gamma,\frac{1}{2}+\gamma \right\vert \left. -\frac{\tilde{x}_-^2}{4} \right)\nonumber \\
  &-&2e^{-i\pi\gamma}\tilde{x}_-^{2-2\gamma} \Gamma(2\gamma-2) {}_{1-\gamma}F_{1-\gamma}\left( 1,\frac{3}{2}-\gamma,2-\gamma \right\vert \left. -\frac{\tilde{x}_-^2}{4} \right). \label{eq:chi-tilde-exact}
  \end{eqnarray}
  This expression has been cross-checked against direct numerical evaluation of the Eq.~(\ref{eq:chi-noeps}). At $|\tilde{x}_-|\gg 1$, Eq.~(\ref{eq:chi-tilde-exact}) has the asymptotics:
  \begin{equation}
  \widetilde{\chi}(\gamma,\tilde{x}_-)=-2\gamma_E-i\pi-2\ln\tilde{x}_- + \tilde{x}_-^{-2\gamma}e^{-i\tilde{x}_-}\frac{2\cos(\pi\gamma) \Gamma(2\gamma)\Gamma\left(\frac{1}{2}-\gamma \right)}{\sqrt{\pi}\Gamma(1-\gamma)}+O(\tilde{x}_-^{-1}). \label{eq:chi-tilde-asy}
  \end{equation}
From the point of view of integral representation (\ref{eq:chi-tilde-Mellin}), the non power-supressed part of the asymptotics (\ref{eq:chi-tilde-asy}) arises due to the pole at $\lambda=0$ to the left of the contour $C_\lambda$, while the  $O(\tilde{x}_-^{-2\gamma})$-suppressed term is due to contribution of essential singularity at $\lambda=\infty$. The essential singularity contributes, because of the $\exp\left[ i\pi \lambda/2 \right]$-factor in Eq. (\ref{eq:delta-chi-gamma-lambda}), which cancels exponential damping  due to $\Gamma$-functions in the quadrant  ${\rm Im}\lambda <0$ and  ${\rm Re}\lambda <0$. Strictly-speaking, this behavior of $\Delta\chi(\gamma,\lambda)$ at $\lambda\to
\infty$ prevents one form closing of the contour $C_\lambda$ around poles to the left of it, hence representation (\ref{eq:chi-tilde-exact}) has been obtained by closing of the contour around right-hand poles, which is allowed by Jordan's lemma. However, as one can see from Eq.~(\ref{eq:chi-tilde-asy}), the contribution of pole at $\lambda=0$ have correctly reproduced non power-suppressed terms in the asymptotics of Eq.~(\ref{eq:chi-tilde-exact}), and therefore our calculation after Eq.~(\ref{eq:BFKL-chi-tilde}) is correct.  

\begin{figure}
\begin{center}
\includegraphics[width=0.45\textwidth]{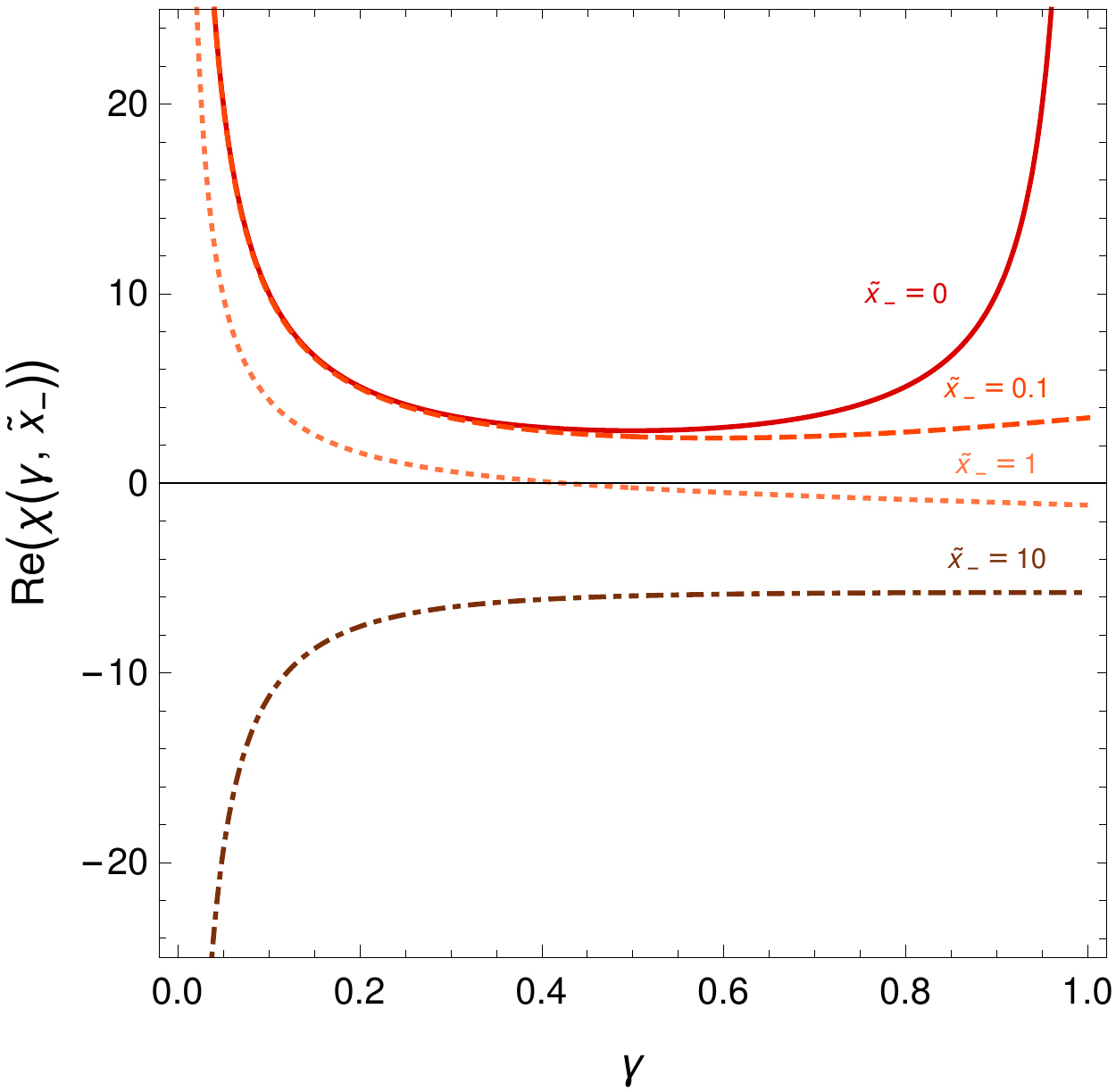}
\includegraphics[width=0.45\textwidth]{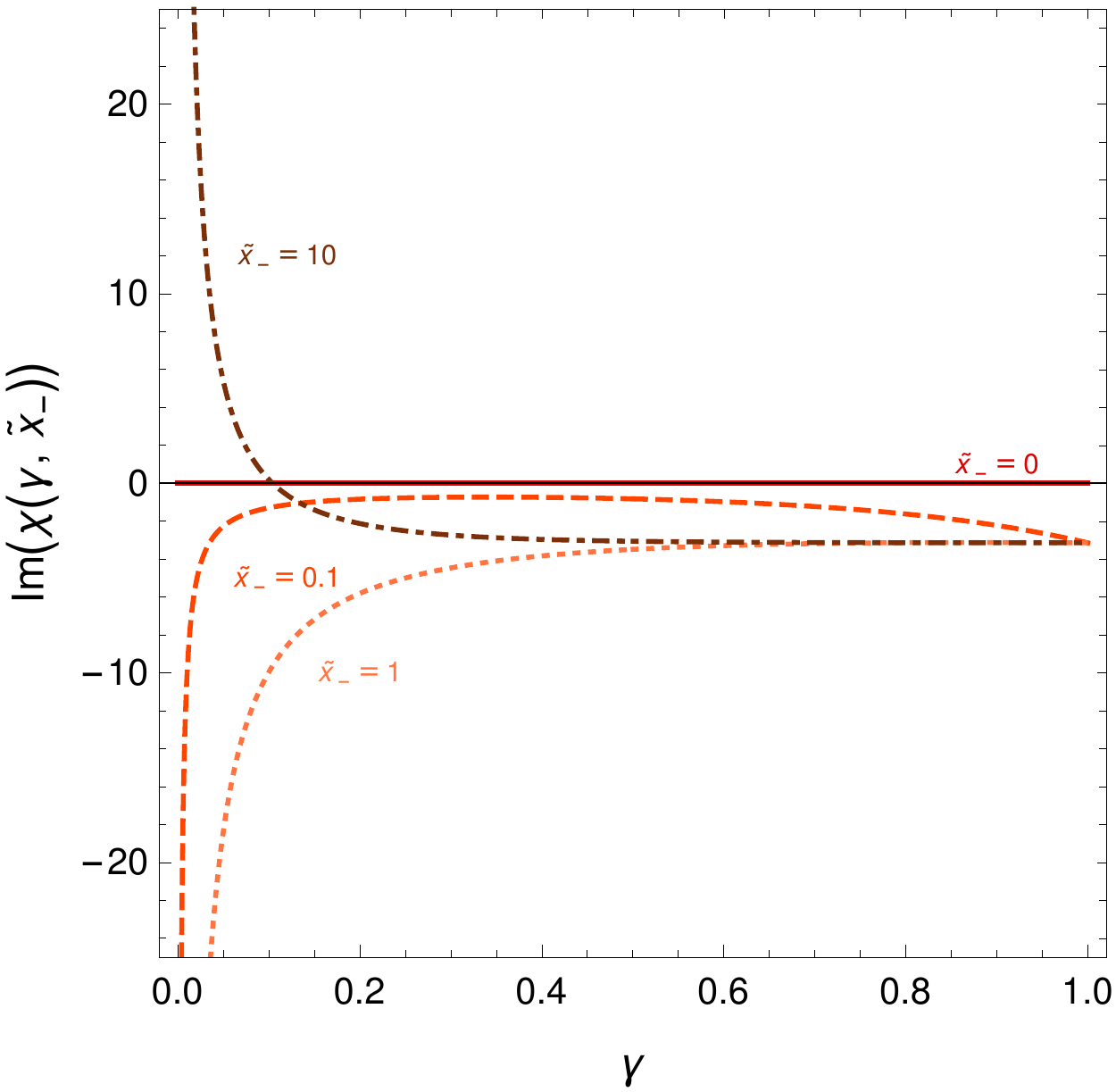}
\end{center}
\caption{The $\gamma$-dependence of real (left-panel) and imaginary (right panel) parts of the $\tilde{x}_-$-dependent characteristic function (\ref{eq:chi-tilde-DEF}) for $\tilde{x}_-=0$ (solid line), $0.1$ (dashed line), $1$ (dotted line) and $10$ (dash-dotted line). \label{fig:chi-gamma}}
\end{figure}

\begin{figure}
\begin{center}
\includegraphics[width=0.45\textwidth]{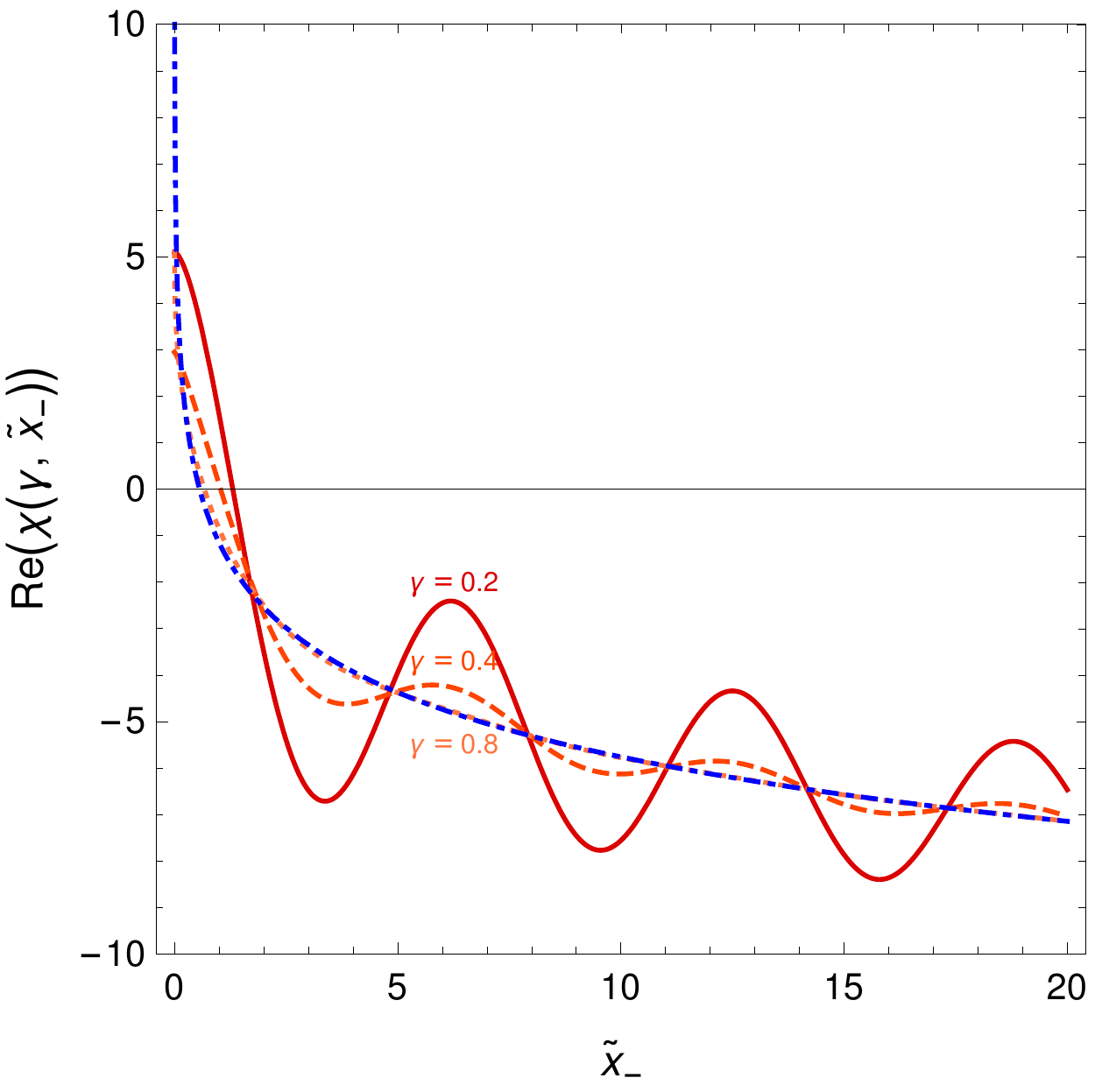}
\includegraphics[width=0.45\textwidth]{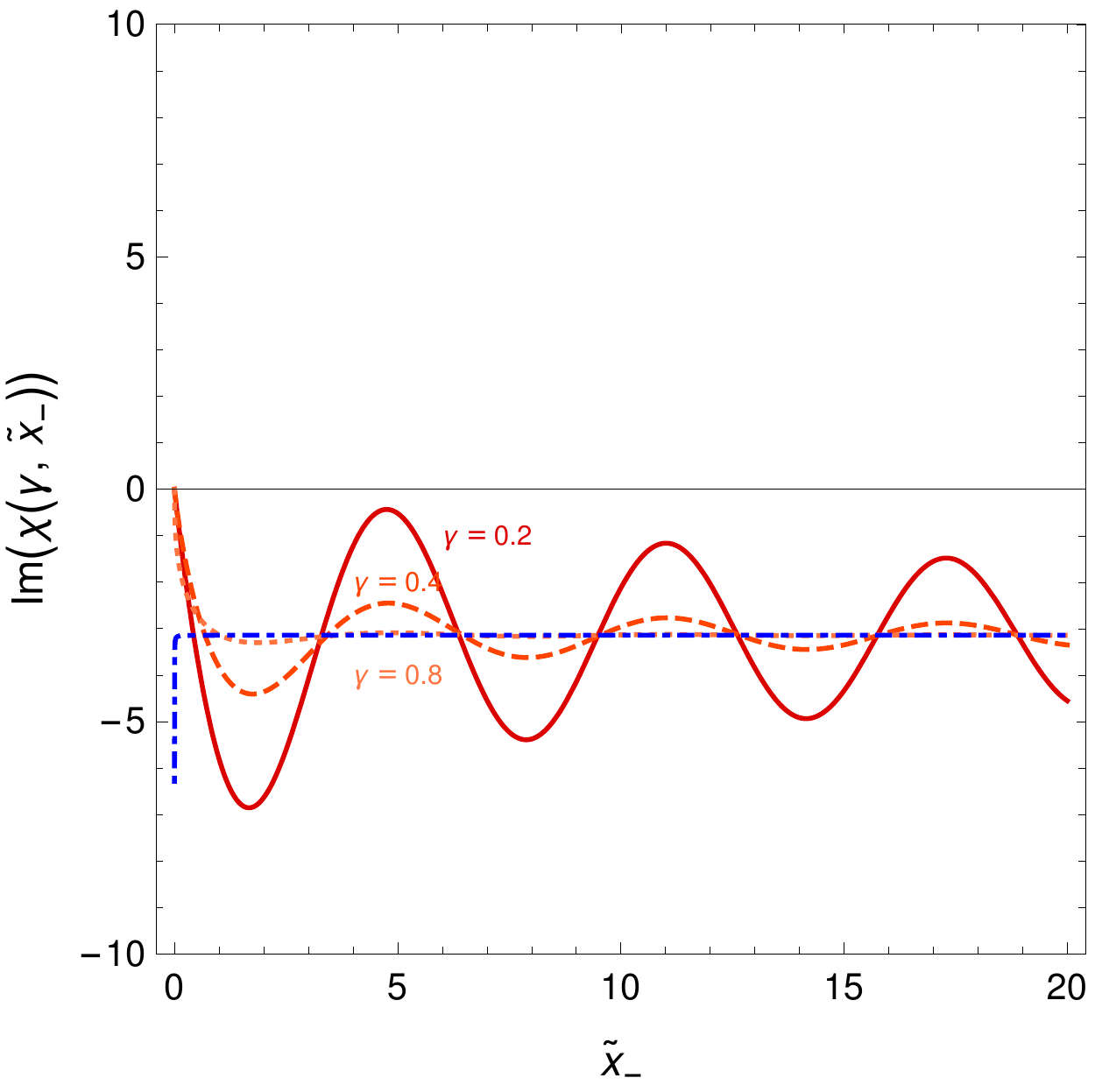}
\end{center}
\caption{The $\tilde{x}_-$-dependence of real (left-panel) and imaginary (right panel) parts of the characteristic function (\ref{eq:chi-tilde-DEF}) for $\gamma=0.2$ (solid line), $0.4$ (dashed line), $0.8$ (dotted line). The non power-suppressed part of asymptotic expression (\ref{eq:chi-tilde-asy}) is shown by the dash-dotted line. \label{fig:chi-x}}
\end{figure}
 
 The modified characteristic function $\widetilde{\chi}(\gamma,\tilde{x}_-)$ is plotted as a function of $\gamma$ in the Fig.~\ref{fig:chi-gamma}. It reduces to Lipatov's one (\ref{eq:Lipatov-chi}) in the limit $\tilde{x}_-\to 0$, but the so-called ``anti-collinear'' pole at $\gamma=1$ is very sensitive to $\tilde{x}_-$ and disappears for any non-zero value of it. The ``collinear'' pole at $\gamma=0$ stays intact, which is to be expected, because it is important for proper factorization of collinear divergences from UPDF. The $\tilde{x}_-$-dependence is shown in the Fig.~\ref{fig:chi-x} for $\tilde{x}_->0$. Dependence for $\tilde{x}_-<0$ follows from the fact, that for real $\tilde{x}_-$ and $\gamma$ one has:
\[
\widetilde{\chi}(\gamma,-\tilde{x}_-)=\widetilde{\chi}^{*}(\gamma,\tilde{x}_-),
\]  
due to definition (\ref{eq:chi-tilde-DEF}). The asymptotics (\ref{eq:chi-tilde-asy}) is satisfied, and we emphasize, that the LL Sudakov formfactor follows essentially from the presence of $\ln 
\tilde{x}_-$-term in this asymptotic expression.  

\bibliography{mybibfile}

\end{document}